\documentclass[twoside,11pt]{article}
\usepackage{amssymb}
\usepackage{latexsym}
\usepackage{amsmath}
\usepackage{xcolor}
\usepackage{lineno}
\usepackage{eurosym}
\usepackage{physics}
\usepackage{todonotes}
\usepackage{booktabs}
\usepackage{multirow}
\usepackage{sectsty}
\usepackage{graphicx}
\usepackage{amsmath}
\usepackage{amssymb}
\usepackage{subcaption}
\usepackage{wrapfig}
\usepackage{caption}
\usepackage{hyperref}
\hypersetup{colorlinks=true,linkcolor=blue,urlcolor=blue}
\usepackage{algorithm}
\usepackage{algpseudocode}
\usepackage{cleveref}
\algnewcommand\MyAnd{\textbf{and} }
\usepackage{float}
\newfloat{algorithm}{t}{lop}
\usepackage{listings}
\definecolor{dkgreen}{rgb}{0,0.6,0}
\definecolor{gray}{rgb}{0.5,0.5,0.5}
\definecolor{mauve}{rgb}{0.58,0,0.82}

\usepackage{fancyhdr}
\usepackage[left=2cm,top=1.5cm,bottom=1.5cm,right=2cm,includefoot,includehead,headheight=13.6pt]{geometry}
\usepackage{lastpage}
\usepackage{enumitem}
\usepackage{lipsum}
\usepackage{titlesec}
\usepackage[toc]{multitoc}

\usepackage{pdflscape}
\usepackage{bm}
\numberwithin{equation}{section}

\newcommand{\ssfill}{\xleaders\hbox to 0.35em{\scriptsize.}\hfill}

\newcommand*{\cventry}[7][.25em]{
  \noindent\begin{tabular*}{\textwidth}{l@{\extracolsep{\fill}}r}%
	  {\bfseries #4} & {\bfseries #5} \%
	  {\itshape #3\ifthenelse{\equal{#6}{}}{}{, #6}} & {\itshape #2}\%
  \end{tabular*}%
  \ifx&#7&%
    \else{\\vbox{\small#7}}\fi%
  \par\addvspace{#1}}

\newcommand*{\hintfont}{\bfseries}
\newcommand*{\hintstyle}[1]{{\noindent\hintfont{#1}}}
\newcommand*{\cvitem}[3][.25em]{%
  \ifthenelse{\equal{#2}{}}{}{\hintstyle{#2}: }{#3}%
  \par\addvspace{#1}}

\let\OriginalQuotation\quotation
\renewcommand*{\quotation}{\OriginalQuotation\small\sf}

\makeatletter
\g@addto@macro\bfseries{\boldmath}
\makeatother
\begin{document}
\allsectionsfont{\sffamily}

\fancyhead{}
\fancyfoot{}

\fancyhead[CO]{{}}
\fancyhead[LO]{{}}
\fancyhead[RO]{{}}
\fancyfoot[R]{\thepage\, / {\color[rgb]{0.6,0.,0}\pageref{LastPage}}}
\renewcommand{\headrulewidth}{0pt}

\newcommand*{\titleGP}{\begingroup 
\thispagestyle{empty}

{\centering 

\begin{center}
{\LARGE {\sf Stochastic automatic differentiation for Monte Carlo processes}} \\[0.2\baselineskip] 
\end{center}

}
\begin{center}
  {\large Guilherme~Catumba, Alberto~Ramos and Bryan~Zaldivar} 
\end{center}
\vspace{0.2cm}
\begin{center}
  {IFIC (CSIC-UVEG). Edificio Institutos Investigación,
       Apt. 22085, E-46071. Valencia, Spain.}
\end{center}

\vspace{1cm}
\begin{center}
  \large{\sf Abstract}
\end{center}
\rule{\textwidth}{0.4pt}
\noindent

Monte Carlo methods represent a cornerstone of computer science. 
They allow to sample high dimensional distribution functions in an
efficient way. 
In this paper we consider the extension of Automatic Differentiation (AD)
techniques to Monte Carlo process, addressing the  
problem of obtaining derivatives (and in general, the Taylor series) 
of expectation values. 
Borrowing ideas from the lattice field theory community, we examine two approaches. 
One is based on reweighting while the other represents an extension of 
the Hamiltonian approach typically used by the Hybrid Monte Carlo
(HMC) and similar algorithms. We show that the Hamiltonian approach
can be understood as a change of variables of the reweighting
approach, resulting in much reduced variances of the coefficients of
the Taylor series. 
This work opens the door to find other variance reduction techniques
for derivatives of expectation values.


\noindent\rule{\textwidth}{0.4pt}\\[\baselineskip] 

\tableofcontents

\newpage
\endgroup}

\begingroup 
\thispagestyle{empty}

{\centering 

\begin{center}
{\LARGE {\sf Stochastic automatic differentiation for Monte Carlo processes}} \\[0.2\baselineskip] 
\end{center}

}
\begin{center}
  {\large Guilherme~Catumba, Alberto~Ramos and Bryan~Zaldivar} 
\end{center}
\vspace{0.2cm}
\begin{center}
  {IFIC (CSIC-UVEG). Edificio Institutos Investigación,
       Apt. 22085, E-46071. Valencia, Spain.}
\end{center}

\vspace{1cm}
\begin{center}
  \large{\sf Abstract}
\end{center}
\rule{\textwidth}{0.4pt}
\noindent

Monte Carlo methods represent a cornerstone of computer science. 
They allow to sample high dimensional distribution functions in an
efficient way. 
In this paper we consider the extension of Automatic Differentiation (AD)
techniques to Monte Carlo process, addressing the  
problem of obtaining derivatives (and in general, the Taylor series) 
of expectation values. 
Borrowing ideas from the lattice field theory community, we examine two approaches. 
One is based on reweighting while the other represents an extension of 
the Hamiltonian approach typically used by the Hybrid Monte Carlo
(HMC) and similar algorithms. We show that the Hamiltonian approach
can be understood as a change of variables of the reweighting
approach, resulting in much reduced variances of the coefficients of
the Taylor series. 
This work opens the door to find other variance reduction techniques
for derivatives of expectation values.


\noindent\rule{\textwidth}{0.4pt}\\[\baselineskip] 

\tableofcontents

\newpage
\endgroup

\section{Introduction}
\label{sec:intro}

Monte Carlo (MC) techniques are ubiquitous in science, and
particularly in physics. From particle physics to cosmology, in order
to extract information about the quantities (parameters, quantum
fields, etc) of our physics models we mostly rely on MC, since it
provides an unbiased estimator of the underlying probability
distributions of such quantities, and consequently, of expectation
values of our observables: 
\begin{equation}
  \label{eq:exp_cost}
  \mathbb E_{p_{\theta}} [f(x; \theta)]\,.
\end{equation}
Here $p_{\theta}(x)$ is the distribution of the quantities of interest $x$,  and we
will consider the case where it also depends on the 
parameters $\theta$ (both $x$ and $\theta$ could be multivariate). Two
prominent examples arise in physics: {\it i)} in the context of a
quantum field theory, $x$ represent the quantum fields, while $\theta$
could represent for example the mass of the fields, as well as their
couplings; {\it ii)} in the context of fitting physics models to some
observed data,  $x$ represent the model parameters (e.g. in a
cosmological model, the matter energy density, the Hubble parameter,
etc), while $\theta$ are the so-called
\emph{hyper-parameters}\footnote{These are parameters which the
  analysist considers as deterministic, so they do not follow a
  distribution themselves, but do determine the distribution of the
  ``standard'' parameters.} that fix either their prior distribution,
or the likelihood of the data, or both, for example in a Bayesian
inference framework.


 In many cases we are interested in optimizing such expectations w.r.t. $\theta$. The state-of-the-art in optimization is represented by Stochastic Gradient Descent (SGD) methods, especially popular in the statistics and machine learning communities. For the cases at hand, the application of any variant of the SGD algorithm requires
to determine the gradient of the expectation cost
Eq.~(\ref{eq:exp_cost}) w.r.t  $\theta$. 

In the case of Bayesian inference, one of the interesting questions arises when we are concerned about the sensitivity of the Bayesian predictions on the hyper-parameters. Formally these predictions\footnote{More concretely, the distribution of such predictions, known as the {\it predictive distribution} in the Bayesian jargon.} are determined as expected values of the form given in Eq.~(\ref{eq:exp_cost}), and we are interested in
the dependence w.r.t. $\theta$. It is important to note at this point that nowadays, in the Bayesian inference community, the above question is -to our knowledge- not addressed. Indeed, the Bayesian predictions are commonly calculated at optimal values $\boldsymbol{\theta}_{\rm opt}$ of the hyper-parameters, the latter being obtained from a point-wise Maximum Likelihood optimization of an approximation of the Bayesian evidence, or with ``Bayesian optimization'' methods. Typically no further analysis is performed quantifying the impact on the predictions when the hyper-parameters values deviate from $\theta_{\rm opt}$.

Another situation in Bayesian inference when the above problem appears is actually very common, specifically in the context of approximate inference, where the aim is to approximate the true posterior $p_{\theta}(x)$ by a distribution $q_{\phi}(x)$. Popular implementations minimize either the forward Kullback-Leibler divergence, KL$[p_{\theta}||q_{\phi}]$, or its reverse: KL$[q_{\phi}||p_{\theta}]$. Since the posterior is unknown (it is what such methods try to approximate in first place), the forward KL can be estimated via reweighting: FKL = $(1/Z_w) E_{q_{\phi}}[w \log(\hat{p}_{\theta}/q_{\phi})]$ (see e.g. \cite{Jordan2021}), where $w=\hat{p}_{\theta}/q_{\phi}$, and $\hat{p}_{\theta}$ is the unnormalized posterior (assumed to be tractable), and $Z_w = E_{q_{\phi}}[w]$. In this case one is interested in minimizing the FKL w.r.t. the parameters $\phi$. In cases where the objective function is the reverse KL (typical case in Variational Infererence) the expectation is directly with respect to $q_{\phi}$. While for simple choices of the latter the procedure is well defined (see below Sect.\ref{sec:existing}), in general it is a complex task when we want to minimize w.r.t. parameters which are implicit in the samples.

In all these cases the key question is the determination of the
gradient and (possibly) higher derivatives 
\begin{equation}
  \label{eq:gradients}
  \frac{\partial}{\partial \theta_j }\mathbb E_{p_{\theta}}[f(x; \theta)]\,,\quad
  \frac{\partial^2}{\partial \theta_j\partial \theta_k }\mathbb E_{p_{\theta}}[f(x; \theta)]\,,\dots
\end{equation}
of expected values, where $\theta_j$ are the components of $\theta$.

In this work we focus on the typical case when the relevant expectations
values Eq.~(\ref{eq:exp_cost}) are determined using some Monte Carlo
(MC) method:
\begin{equation}
  \mathbb E_{p_{\theta}}[f(x; \theta)] \approx
  \frac{1}{N_s}\sum_{s=1}^{N_s} f(x_s; \theta)~,
  \label{eq:MC_est}
\end{equation}
where $x_s$ are samples from the unnormalized $\hat p_{\theta}$. \emph{Our aim is to develop a formalism to compute the gradients of expr.(\ref{eq:MC_est}) with respect to $\theta$, using automatic differentiation techniques.}

\subsection{Existing methods}
\label{sec:existing}
There are solutions in the literature for this problem: for some simple distributions the reparametrization trick can be used, which is nothing but the ability to express a sample $x$ as a deterministic function $g({\theta,\eta})$, of the parameters $\theta$ and a random variable $\eta$ \emph{that does not depend on $\theta$}. A typical example is when $p_\theta$ is a multivariate Gaussian with mean ${\mu}(\theta)$ and covariance matrix ${\bf C}$, in whose case a sample $x_s$ can be expressed as $x_s = {\mu}(\theta) + {\bf L}\cdot{\eta}_s$, where ${\bf L}$ is the Cholesky decomposition of ${\bf C}$, and the sampled random variable ${\eta}_s\sim{\cal N}({\bf 0},\mathbb{I})$. Clearly, since $x_s$ is an explicit function of $\theta$, so it will be $f(x_s,\theta)$ in Eq.(\ref{eq:MC_est}), making possible to use the usual techniques of Automatic Differentiation to obtain the gradients w.r.t. $\theta$ exactly. For other popular distributions as Gamma, Beta or Dirichet, among others, this simple reparametrization is not possible, and generalizations of the above trick have been developed (e.g. in \cite{Figurnov2018}, using implicit differentiation). Nonetheless, the considered distributions should still be somehow reparametrizable. 


Another existing alternative is to use the \emph{score function estimator}, allowing us to obtain the gradient for a more general case. This method just uses the trivial relation
\begin{equation}
  \nabla_{\theta} \log p_{\theta}(x) = \frac{1}{p_{\theta}(x)}
  \nabla_{\theta} p_{\theta}(x) \,,
\end{equation}
such that the gradient in expr.(\ref{eq:gradients}) could be approximated with MC as:
\begin{equation}
  \nabla_{\theta} \mathbb E_{p_{\theta}}[f(x; \theta)]
  \approx
  \frac{1}{N_s}\sum_{s=1}^{N_s}
  \Big[
    (\nabla_{\theta} \log p_{\theta}(x_s)) f(x_s;\theta)
    + \nabla_\theta f(x_s;\theta)
    \Big]
    \label{eq:sfe}
\end{equation}

While certainly being a more flexible method, this estimator is known to suffer in practice from large sample-to-sample variance
(although see \cite{Lievin2020} for variance-control methods in this context). Lastly, other treatments have been proposed which attempt to extract the best of both methods above, i.e. to be applicable to distributions beyond those typically reparametrizable, while keeping a low variance (\cite{cong2018go}). 

Up to our knowledge, all the existing efforts applying the solutions mentioned above require to know the normalization of the
distribution $p_{\theta}(x)$, which prevents the use in conjunction with
Monte Carlo methods, where samples of a distribution are often
obtained in the case where the corresponding normalization is unknown.

The question on how to determine derivatives of expected values taken
over complicated distributions $p_\theta(x)$, especially in the
case that one relies on Monte Carlo methods to draw samples from such a distribution is still open. 
Ideally one would like an ``automatic'' procedure, i.e. 
extending the benefits of automatic differentiation to Monte Carlo
processes. In this work we will explore two such approaches.
First we use the idea of reweighting, where the expectation values, \cref{eq:MC_est}, are modified by a weighting factor that takes into account the dependence on the parameters, but that utilizes unmodified samples for the average.
The second method is a modification of Hamiltonian sampling algorithms, that includes the generation of samples that carry themselves the information about the parameters.
Both methods can be used for unnormalized probability distributions and allow the computation of derivatives of arbitrary orders.


\section{Automatic differentiation}
\label{sec:ad}

Our methods to compute derivatives of expectation values are based on
the techniques of automatic differentiation (AD). 
By AD we understand a set of techniques to determine the derivative of
a deterministic function specified by a computer program. 
There are various flavors of AD in the market, and our algorithms are
quite agnostic about the particular implementation that is used, but
in order to make the proposal and notation more concrete we are going
to choose a particular method, based on operations of polynomials
truncated at some order. The generalization of our techniques to
other flavors of AD is straightforward. 

In what follows it is useful to use multi-index notation. The
$d$-dimensional multi-index is defined by
\begin{equation}
  n = (n_1,\dots,n_d)\,.
\end{equation}
We can define a partial order for multi-indices by the condition
\begin{equation}
  n \le m \Longleftrightarrow n_i \le m_i\quad \forall i=1,\dots,d\,.
\end{equation}
We also define the absolute value, factorial and power by the relations
\begin{equation}
  |n| = \sum_{i=1}^d n_i\,,;\quad
  n! = \prod_{i=1}^d n_i!\,,\quad
  \epsilon^n = \prod_{i=1}^d \epsilon_i^{n_i}\,.
\end{equation}
Finally the higher-order partial derivative is defined by
\begin{equation}
  \frac{\partial^n}{\partial x^n} = \frac{\partial^{|n|}}{\partial x_1^{n_1}\cdots\partial x_d^{n_d}} \,.
\end{equation}

With this notation, polynomials 
of degree $p = (p_1,\dots,p_d)$ in several variables
$\{\epsilon_i\}_{i=1}^d$ are represented in the compact form
\begin{equation}
  \tilde a(\epsilon) = \sum_{n \le p} c_{n} \epsilon^{n}~.
  \label{eq:truncpol}
\end{equation}
Note that each variable $\epsilon_i$ is raised at most at the power
$p_i$ and that the index of the coefficient $c_n$ is itself a multi-index (i.e. 
$c_n = (c_{n1}, c_{n2},\dots, c_{nd})$). If the coefficients $c_{ni}$
are elements of a field (i.e.  
real numbers), the
addition/multiplication of these polynomials where terms $\mathcal
O(\epsilon_j^{n_j+1})$ are neglected form an algebra over the very same field. 

As an example calculation in this algebra, consider the following two polynomials in
two variables and with degrees $p=(2,3)$
\begin{eqnarray}
  \tilde a(\epsilon) &=& 2 + \epsilon_1 + \epsilon_2^3\,, \\
  \tilde b(\epsilon) &=& 1 + \epsilon_1 + 2\epsilon_1^2 + 3\epsilon_2^2\,,
\end{eqnarray}
we have
\begin{eqnarray}
  (\tilde a+\tilde b)(\epsilon) &=& 3 + 2\epsilon_1 + 2\epsilon_1^2 + 3\epsilon_2^2+ \epsilon_2^3\,, \\
  (\tilde a\cdot \tilde b)(\epsilon) &=& 2 + 3\epsilon_1 + 5\epsilon_1^2 + 2\epsilon_1^3 + 3\epsilon_1\epsilon_2^2 + \epsilon_1\epsilon_2^3 + 2\epsilon_1^2\epsilon_2^3 + \epsilon_2^3\,.
\end{eqnarray}
In particular it is important to note that we have dropped terms
$\propto \epsilon_1^3, \epsilon_2^5$ in the product, since $3 > p_1=2$
and $5>p_2=3$. 
Therefore the ``$=$'' sign in the above equations has to be understood as
``\emph{up to higher order corrections}''.
Elementary operations and functions acting on these polynomials can be
defined in a straightforward way (see~\cite{Haro2022}).

The connection of the algebra of truncated polynomials with AD is a
consequence of Taylor theorem. 
Let $f(x)$ be a deterministic function in the variables $x_1,\dots,x_d$. 
If each variable is promoted to a truncated polynomial
\begin{equation}
  x_i \longrightarrow \tilde x_i(\epsilon) = x_i + \epsilon_i\,,
\end{equation}
and we evaluate the function $f$ with the truncated polynomials as
input
\begin{equation}
  \tilde f(\epsilon) = f(\tilde x(\epsilon)) = \sum_{n\le p} f_n\epsilon^n\,, 
\end{equation}
it is easy to see that the result is a polynomial that is equal to
the Taylor series of $f$ at $x$. 
In particular partial derivatives of the function are obtained by the
relation 
\begin{equation}
  f_n = \frac{1}{n!} \frac{\partial^n f}{\partial x^n}~.
\end{equation}

Note that the analogy with the Taylor expansion is just at the level of the coefficients $f_n$, which are obtained automatically when writing functions of truncated polynomials $\tilde x_i(\epsilon)$, while $\epsilon$ is exclusively a symbolic quantity.


\section{AD for Monte Carlo process}
\label{sec:mth}

\subsection{Reweighting and Automatic Differentiation}
\label{sec:reweighting}

Samples $\{x^\alpha\}_{\alpha=1}^N$ of some distribution
$p_{\theta}(x)$ allows to estimate expectation values
\begin{equation}
  \mathbb E_{p_\theta}[f(x)] = \frac{1}{N}\sum_{A=1}^N f(x^\alpha;\theta) + \mathcal
  O\left( \frac{1}{\sqrt{N}} \right) \,.
\end{equation}
In this expression, $\theta$ are some parameters that the distribution
function (and possibly the function $f$) depends on. 
We are interested in obtaining the gradient of expectation values with
respect to the parameters $\theta$
\begin{equation}
  O_n = \frac{\partial^n}{\partial \theta^n} \mathbb E_{p_\theta}[f(x; \theta)]\,.
\end{equation}

Our proposal to determine these derivatives is based on
\emph{reweighting} (a.k.a. importance sampling). 
If we have $N$ samples $\{x^\alpha\}_{\alpha=1}^{N}$ of the
distribution $p_\theta$ they can be   
used to determine expectation values of a
different distribution $p'$ thanks to the identity
\begin{equation}
  \label{eq:rw}
  \mathbb E_{p'}[f(x;\theta)] = \frac{\mathbb E_{p_\theta}\left[\frac{p'}{p_\theta} f(x;\theta)\right]}{\mathbb E_{p_\theta}\left[ \frac{p'}{p_\theta} \right]} = 
  \frac{\sum_{\alpha=1}^N w^\alpha f(x^\alpha;\theta)}{\sum_{\alpha=1}^N w^\alpha} + \mathcal
  O\left( \frac{1}{\sqrt{N}} \right)\,,
\end{equation}
where $w^\alpha = p'(x^\alpha)/p_\theta(x^\alpha)$ are usually called
\emph{reweighting factors}. 
Our approach consists in using for the target distribution
\begin{equation}
  p'(x) = p_{\tilde \theta(\epsilon)}(x)\,, \qquad (\tilde \theta_i(\epsilon) = \theta_i + \epsilon_i)\,.
\end{equation}
With this substitution, the reweighting factors will become truncated
polynomials
\begin{equation}
  \tilde w^\alpha(\epsilon) = \frac{p_{\tilde \theta}(x^\alpha)}{p_\theta(x^\alpha)} =
  \sum_{n \le p}w^\alpha_{n}\epsilon^n,
\end{equation}
with leading coefficients $w^{\alpha}_0 = 1$. 
The basic Eq.~(\ref{eq:rw}) now leads to estimates for the expectation
values in the form of truncated polynomials
\begin{equation}
  \label{eq:rwformula}
  \frac{\sum_{\alpha=1}^N \tilde w^\alpha f(x^\alpha;\tilde \theta)}{\sum_{\alpha=1}^N \tilde w^\alpha} = 
  \sum_{n\le p} O_n\epsilon^n\,,  
\end{equation}
that will give stochastic estimates the Taylor series coefficients of
the expectation values.  i.e.
\begin{equation}
  O_n = \frac{1}{n!} \frac{\partial ^n}{\partial \theta^n} E_{p_\theta}[f(x; \theta)] + \mathcal O \left( \frac{1}{\sqrt{N}} \right)\,.
\end{equation}
Borrowing the terminology of the lattice field theory community, we
distinguish two type of contributions to the derivatives in
Eq.~(\ref{eq:rwformula}):
\begin{description}
\item[Connected contributions:] They come from the explicit dependence
  of the observable $f(x;\tilde \theta)$ on the parameters $\theta$.
\item[Disconnected contributions:] They come from the reweighing
  factors $\tilde w^\alpha$, and account for the implicit dependence
  that the samples $x^\alpha$ have on the parameters $\theta$.
\end{description}

It is important to note that the denominator in Eq.~(\ref{eq:rw})
accounts for the possibility that the distributions are 
unnormalized (i.e this expression is valid in the context of Monte
Carlo sampling, where samples are obtained without knowledge of the
normalization of $p_\theta$).

This reweighting approach can be thought as a generalization of the
score function estimator described in sec.\ref{sec:intro}: indeed, in
the case of normalized distributions, and if one considers only
the first derivatives, the reweighting factors become $\tilde w^\alpha = 1 +
(\nabla_\theta\log p_\theta)\epsilon$, with the first-order coefficient
being precisely the term appearing in expr.(\ref{eq:sfe}).


\subsection{Hamiltonian perturbative expansion}
\label{sec:nspt}

Sampling algorithms based on Hamiltonian dynamics are nowadays a
central tool in many different areas. 
Probably the best known example is the Hybrid Monte Carlo (HMC)
algorithm.
Originally developed in the context of Lattice
QCD~\cite{DUANE1987216}, today it is also a cornerstone in Bayesian inference.

The HMC algorithm belongs to the class of Metropolis-Hastings
algorithms that allows to obtain samples of arbitrarily complex
distribution functions with a high acceptance rate.  In order to sample
the distribution function
\begin{equation}
  p_\theta(x) = \frac{1}{\mathcal Z}\exp \left\{ -S(x;\theta) \right\}\,, \qquad \left(  \mathcal Z = \int {\rm d} x\, e^{-S(x:\theta)} \right) \,,
\end{equation}
where $x^\alpha\in \mathbb R^d$, we introduce some momentum variables
$\pi^\alpha$, conjugate to $x^\alpha$, and 
consider the sampling of the modified distribution function
\begin{align}
  q_\theta(\pi,x) = \frac{1}{\mathcal Z'} \exp \left\{ -H(\pi, x;\theta) \right\}\,, && \left(  \mathcal Z' = \int {\rm d} x {\rm d} \pi\, e^{-H(\pi,x:\theta)} \right)\,.
\end{align}
Assuming that the momenta are distributed as a standard Gaussian, the Hamiltonian is defined by
\begin{equation}
  H(\pi,x;\theta) = \frac{1}{2}\sum_{\alpha=1}^d\pi^\alpha\pi^\alpha + S(x;\theta)\,.
\end{equation}

\begin{wrapfigure}{l}{0.4\textwidth}
\centering
    \includegraphics[width=0.35\textwidth]{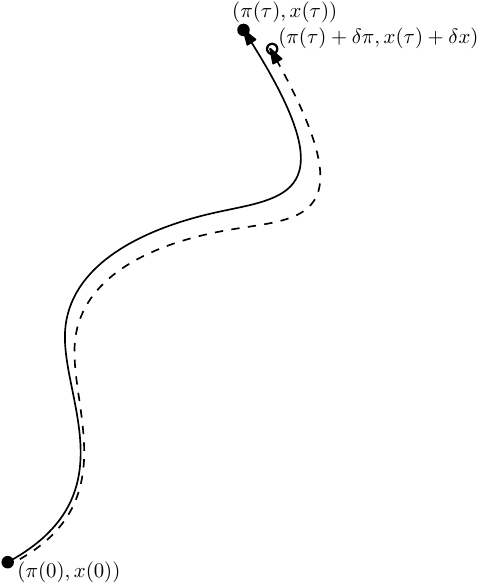}
    \caption{The HMC algorithm makes a new sample proposal $x(\tau)$
      from a previous sample $x(0)$ of the
      target distribution $p_\theta(x)$ by numerically integrating the equations of
      motion of a fictitious Hamiltonian. 
      AD techniques can be applied to determine the dependence of the
      trajectory on model parameters $\theta$.} 
    \label{fig:hmc}
\end{wrapfigure}

It is clear that expectation values of quantities that depend only on
the variables $x$ are the same if they are computed using
$p_\theta(x)$ or $q_\theta(\pi,x)$ (i.e. 
$\mathbb E_{p_\theta}[f(x)] = \mathbb E_{q_\theta}[f(x)]$). 
On the other hand the distribution $q_\theta(\pi,x)$  can be
sampled with a high acceptance rate just by 1) throwing randomly
distributed Gaussian momenta $\pi(0)\sim e^{-\pi^2/2}$, 2) solving the
Hamilton equations of motion (eom)
\begin{eqnarray}
  \label{eq:eom}
  \dot x^\alpha &=& \pi^\alpha\,, \\
  \dot \pi^\alpha &=& - \frac{\partial H}{\partial x^\alpha} = - \frac{\partial S}{\partial x^\alpha}\,,
\end{eqnarray}
for a time interval from $(\pi(0),x(0))$ to $(\pi(\tau),x(\tau)$, and
3) performing a Metropolis-Hastings accept/reject step with probability 
$e^{-\Delta H}$ (see Figure \ref{fig:hmc}). The values of $x(\tau)$ are distributed according to
the probability density $p_\theta(x)$ (for a
proof see the original reference~\cite{DUANE1987216}). The trajectory
length $\tau$ can be chosen 
arbitrarely, although in order to guarantee the ergodicity of the
algorithm it is required that this length is chosen randomly from some
distribution (exponential or uniform are the most common choices). 
This point is usually not relevant, but for the case of ``free''
theories (i.e. 
Gaussian distributions $p_\theta$), it is well known that a constant trajectory
length can lead to wrong results (see~\cite{RHMC2017}).

In the last step $\Delta H = H(\pi(\tau),x(\tau);\theta) -
H(\pi(0),x(0);\theta)$ is just the energy violation. 
Since energy is conserved in Hamiltonian systems, this violation of
energy conservation is entirely due to the
fact that the eom in eq.~(\ref{eq:eom}) are solved
numerically, and not exactly. 
Nevertheless this integration of the eom can be made very precise with
a modest computational effort, allowing to reach arbitrarily high
acceptance rates.

The Stochastic Molecular Dynamics (SMD) algorithm is closely related with the HMC, and also based on a
Hamiltonian approach. 
In this case, after each integration step (from $t$ to
$t+\delta t$) of the equations of motion, 
the momenta is partially refreshed according to the equations
\begin{equation}
  \pi \to c_1\pi + \sqrt{1-c_1^2}\, \eta\,, 
\end{equation}
where $\eta $ is a new random momenta with Gaussian distribution, and
$c_1 = e^{-\gamma \delta t}$ is a parameter that can be chosen
arbitrarily via the value of $\gamma$. 
The SMD algorithm has some advantages from the theoretical point of
view, specially in the context of simulating field theories on the
lattice~\cite{Luscher:2011qa}

\subsubsection{Implementation of AD and convergence}
\label{sec:impl-conv}

The typical implementation of either the HMC or the SMD algorithm
involves the numerical integration of the eqs.~(\ref{eq:eom}).
This is performed by using a sequence of steps
\begin{eqnarray}
\label{eq:schemes}    
  \mathcal I_{\pi,h}: \quad \pi \to \pi - h \frac{\partial S}{\partial x}\,\\     
  \mathcal I_{x,h}: \quad x \to x + h\pi\nonumber \,.
\end{eqnarray}
An example is the well known \emph{leapfrog} integrator, that is
obtained by applying the series of steps $\mathcal
I_{\pi,\delta t/2}\mathcal I_{x,\delta t}\mathcal
I_{\pi,\delta t/2}$. Better precision can be obtained by using higher
order schemes (see~\cite{OMELYAN2003272}).

The application of AD to solve the eom eq.~(\ref{eq:eom}) follows
basically the same procedure, with the difference that 
\begin{enumerate}
\item Both coordinates and conjugate momenta variables are promoted to
  truncated polynomials
  \begin{eqnarray}
    x^\alpha &\longrightarrow& \tilde x^\alpha = \sum_i x_{i}^\alpha\epsilon^i\,, \\
    \pi^\alpha &\longrightarrow& \tilde \pi^\alpha = \sum_i \pi_{i}^\alpha\epsilon^i\,.
  \end{eqnarray}
  At the same time the model parameters are also promoted using
  $\tilde \theta_i = \theta_i + \epsilon_i$.  Only the lowest order
  $\pi^\alpha_0$ is initially set with Gaussian momenta
  \begin{equation}
    \tilde\pi^\alpha_0(0) \sim e^{-\pi^\alpha_0 \pi^\alpha_0/2}\,,
  \end{equation}
  while higher orders are initialized to zero.

\item The eom eq.~(\ref{eq:eom}) are solved consistently (i.e.
  order by order in $\epsilon$). Note that the non trivial eom can be
  written at each order as
  \begin{equation}
    \label{eq:dotpi}
    \dot\pi^\alpha_{n} = - \frac{\partial^2 S}{\partial x^\alpha \partial x^\beta} x^\beta_{n} + \text{lower order terms}\,.
  \end{equation}
  and therefore the eom eq.~(\ref{eq:eom}) can be solved numerically
  using the same basic building blocks defined by eq.~(\ref{eq:schemes}).
\item Since now the energy violation $\Delta H$ is a truncated
  polynomial, the usual accept/reject step cannot be carried out. 
  This means that one has to extrapolate the HMC results to
  zero step size $\delta t \to 0$. 
  In practice it is enough to work at sufficiently small step size
  such that any possible bias is well below our statistical uncertainties~\cite{brida_smd-based_2017}.
  
\end{enumerate}
The generalization of the SMD algorithm follows the same basic rules.

Note that the HMC algorithm, despite being in the class of
Metropolis-Hastings algorithms, is a completely deterministic
algorithm in the limit that the eom are integrated exactly. 
This explains the strategy: we are using the tools of AD to determine the Taylor
expansion of $x(\tau)$ with respect to the model parameters $\theta$. 

This approach to the perturbative sampling is intimately related with
the techniques of Numerical Stochastic Perturbation Theory (NSPT)~\cite{DiRenzo:1994av},
especially the Hamiltonian versions described
in~\cite{brida_newmethod_2017,brida2015numerical}.  The crucial
difference is that NSPT is usually applied to determine the deviations
with respect to the ``free'' (i.e. 
Gaussian) approximation, implementing a numerical approach to lattice
perturbation theory, whereas in our case we determine the dependence
with respect to arbitrary parameters $\theta$ (possibly more than one!) present in the
distribution function $p_\theta(x)$.

In practice the numerical implementation of this procedure is
straightforward, and just amounts to numerically solving the eom using
the algebraic rules of truncated 
polynomials (see section~\ref{sec:ad}).

These steps will result in a series of samples $\{\tilde
x_A\}_{A=1}^{N}$ (which are truncated polynomials, cf. expr.(\ref{eq:truncpol})). 
The usual MC evaluation of expectation values 
\begin{equation}
  \frac{1}{N} \sum_A O(\tilde x_A) = \sum_i \bar O_i\epsilon^i\,,
\end{equation}
will give a truncated polynomial that contains the dependence of
expectation values with respect to the model parameters $\theta$.

The convergence of expectation values is guaranteed if the Hessian
\begin{equation}
  H^{\alpha\beta} = \frac{\partial^2 S}{\partial x^\alpha \partial x^\beta}
  \label{eq:Hessian}
\end{equation}
is positive definite (see~\cite{brida_smd-based_2017}).
This condition is always true in the context of perturbative
applications in lattice field theory, since in this case it is
equivalent to having a stable vacuum.  
However, in models defined with compact variables and for the
case or expansions around arbitrary backgrounds, the
convergence of the process is not always guaranteed. 
An important example of this case are the simulation of Yang-Mills
theories on the lattice, where one would in general expect this
process not to converge.
On the other hand, note that in applications of Bayesian inference,
the convergence condition on the Hessian is guaranteed for unimodal
posteriors.


\section{A comparison between approaches}
\label{sec:hamilt-appr-repar}

We have introduced two methods to determine Taylor series (and consequently, derivatives) of
expectation values. 
In the reweighting based method (section~\ref{sec:reweighting}) the
samples obtained by the Monte Carlo method are corrected by the
reweighting factors, eq.~(\ref{eq:rwformula}), to take into account the
dependence on the parameters $\tilde \theta$. 
On the other hand, the Hamiltonian approach (section~\ref{sec:nspt})
produces samples that automatically carry the dependence on the
parameters $\tilde \theta$. It is instructive to see the relation of
each method with the reparametrization trick. 
In order to get some intuition,  we can examine a very simple toy model. 
Imagine that we are interested in the distribution function
\begin{equation}
  \label{eq:Gd}
  p_\sigma(x) = \frac{1}{\sigma\sqrt{2\pi}} e^{-\frac{x^2}{2\sigma^2}}
\end{equation}
and in how expectation values depend on $\sigma$ around $\sigma=1$. 
Of course, this can be trivially solved using a change of variables.
If $\{x_A\}_{A=1}^{N}$ are samples for $ \sigma=1$ then
\begin{equation}
  y_A = \sigma x_A\,,
\end{equation}
are samples for any other value of sigma. 
In particular one can write the truncated polynomial
\begin{equation}
  \label{eq:changeG}
  \tilde x_A = x_A + x_A\epsilon\,, \quad (\epsilon = \sigma-1)\,.
\end{equation}
and now evaluating any expectation value using $\{\tilde x_A\}$ as samples
will produce its corresponding Taylor series. 
In this sense the change of variables can be seen as a particular
transformation to the samples (from $x_A$ to $\tilde x_A$) such that
expectation values evaluated with these samples $\tilde x_A$ give
automatically Taylor series of observables. 

Now consider the reweighting approach to this simple problem. 
One would define $\tilde \sigma = 1+\epsilon$, and determine the
reweighing factors Eq.~(\ref{eq:rwformula}). They read
\begin{equation}
  \tilde w^\alpha(\epsilon) = e^{- \frac{(x^\alpha)^2}{2} \left[ \frac{1}{(1+\epsilon)^2} - 1 \right]    }
\end{equation}
On the other hand, if one performs the change of variables
Eq.~(\ref{eq:changeG}) \emph{before} applying the reweighting formula,
the reweighting factors are given by
\begin{equation}
  \tilde w^\alpha(\epsilon) = e^{- \frac{(x^\alpha)^2}{2} \left[ \frac{1}{(1+\epsilon)^2} - \frac{1}{(1+\epsilon)^2}  \right]    -\log (1+\epsilon)} = \frac{1}{1+\epsilon}\,.
\end{equation}
Note that these reweighting factors are constant (i.e. 
independent on $x$). 
They cancel from the computation of the any expectation value:
\begin{equation}
  \langle O(x) \rangle \approx \frac{\sum_\alpha \tilde w^\alpha O_\alpha}{\sum_\alpha \tilde w^\alpha}\ = \frac{1}{N}\sum_\alpha O_\alpha\,, \qquad \left( O_\alpha = O(\tilde x_\alpha) \right)\,.
\end{equation}

One can therefore see the reparametrization trick as a particular
application of the general reweighting formula, where the change of
variables leads to constant reweighting factors.

We claim that the Hamiltonian approach is just a method to find this
change of variables for complicated distributions and to any order.  
In order to see how this happens, we need to work out the solution of
the equations of motion for our toy model Eq.~(\ref{eq:Gd}).
They read
\begin{eqnarray}
\ddot x_0 &=& - \frac{x_0}{\sigma^2} \,, \\
\ddot x_1 &=& - \frac{x_1}{\sigma^2} + 2\frac{x_0}{\sigma^3}\,.
\end{eqnarray}
It is clear that the equation for $x_0(t)$ is just the usual
harmonic oscillator, with solution
\begin{equation}
  x_0(t) = x_0(0)\cos \left( \frac{t}{\sigma} \right) +
             \sigma \pi_0(0)\sin \left( \frac{t}{\sigma} \right) \,.
\end{equation}
For the next order we have a driven harmonic oscillator (without
damping term). Note, however, that since the frequency of the driven
force is the same as the natural frequency of the oscillator ($\omega
= 1/\sigma$), we have a resonant phenomena: the amplitude of the
oscillations to first orders will increase with the trajectory length.
\begin{figure}
  \centering
  \includegraphics[width=0.9\textwidth]{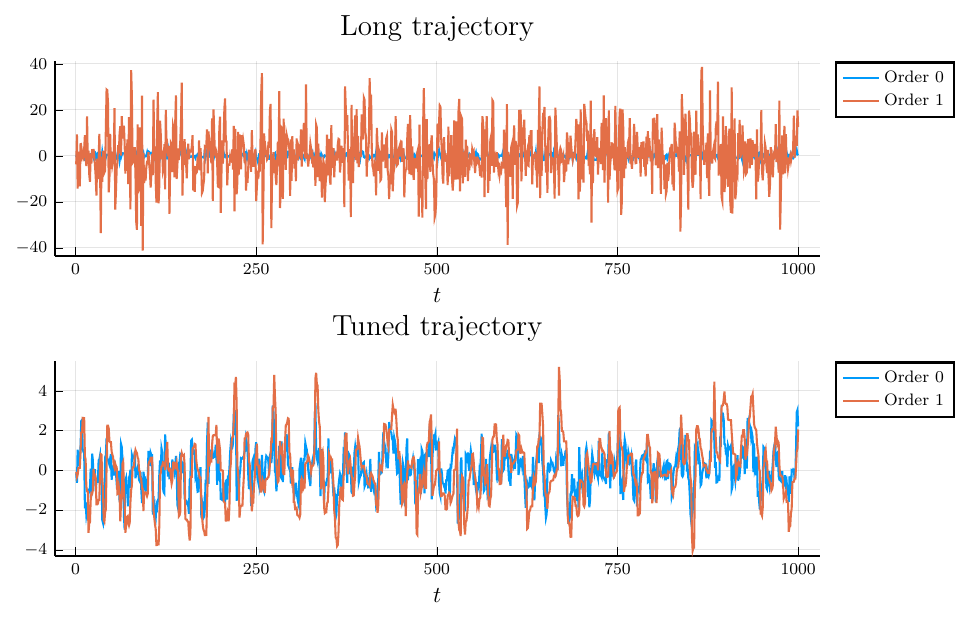}
  \caption{In the Hamiltonian approach the variance of the derivative
    estimators depend crucially on the trajectory length (see text for
    more details).}
  \label{fig:ls}
\end{figure}
It is clear that since the HMC algorithm only integrates the eom
up to a finite time $\tau$, and since we take the average over the samples,
this phenomena does not represent any
issue for the convergence: any trajectory length will produce correct
results according with our expectations. 
On the other hand, it is also clear that the variance of the
observables computed with these samples will depend significantly on
the trajectory length: of the many solutions found by the Hamiltonian
approach (corresponding to different trajectory lengths), some will
produce results with smaller variances. 

Figure~\ref{fig:ls} shows that for certain trajectory length, the
Hamiltonian approach described here just ``finds'' the transformation
given by Eq.~(\ref{eq:changeG}): the zeroth and first order are
very similar. Any other trajectory length will still produce the correct
expectation values, but with a significantly larger variance. 
In the case of the SMD algorithm we observe a similar phenomena, but
in this case it is the parameter $\gamma$ the one that has to be tuned.

This little example shows two important lessons: 1) the Hamiltonian
approach can be considered just a change of variables from the
original samples $x^{\alpha}\rightarrow\tilde y^\alpha(x^\alpha)$ such that once applied to
the reweighting formula eq.~(\ref{eq:rwformula}) gives \emph{constant}
reweighting factors, and 2)
The variance obtained for the derivatives depends on the particular
change of variables. 
In section~\ref{sec:lamphi4} we will comment on the differences in
variance between the two methods in detail.


\section{Some general applications}
\label{sec:applications}

In this section we explore a few applications of the techniques
introduced in section~\ref{sec:mth}. First we consider the application
of the reweighting technique to an optimization problem. 
Second, we consider the application in Bayesian inference to obtain
the dependence of predictions on the parameters that characterize the
prior distribution.

\subsection{Applications in optimization}
\label{sec:opt}

As an example application of an optimization process we will consider
the probability density function
\begin{equation}
  p_\theta(x) = \frac{1}{\mathcal Z}\exp \left\{ -S(x;\theta) \right\}\,, \qquad \left(  \mathcal Z = \int {\rm d} x\, e^{-S(x:\theta)} \right) \,.
\end{equation}
with
\begin{equation}
  S(x; \theta) = \frac{1}{\theta_1^2+1} \left( x_1^2 + x_1^4 \right) + \frac{1}{2}x_2^2 + \theta_2 x_1x_2\,.
\end{equation}

The shape of $S(x;\theta)$ is inspired in the action of a quantum
field theory in zero dimensions, where $x_1$ and $x_2$ are two fields
with coupling $\theta_2$, while $\theta_1$ is related to the mass of
the field $x_1$.
Expectation values with respect to $p_\theta(x)$ are functions of
the parameters $\theta$.  

\begin{figure}
  \centering
  \includegraphics[width=0.8\textwidth]{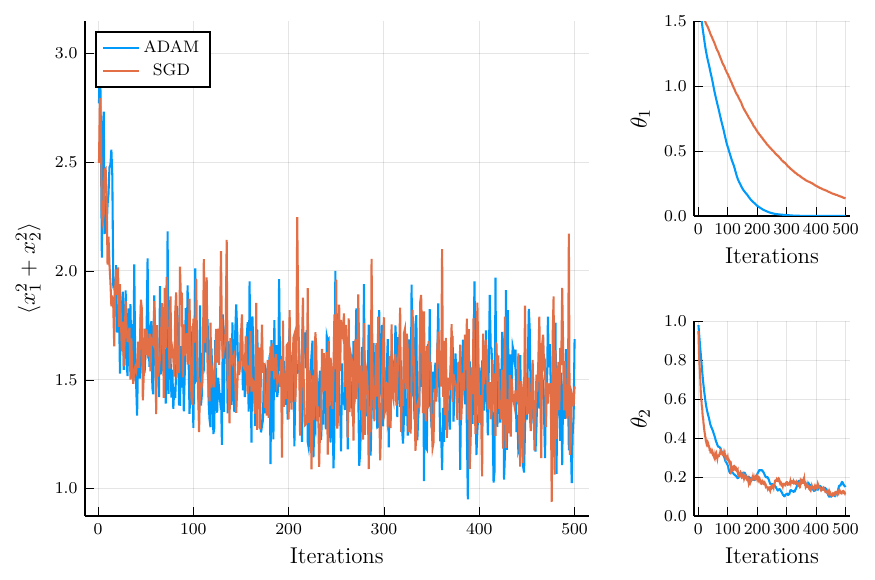}
  \caption{Estimates of the objective function
    $\mathbb E_\theta[x_1^2+x_2^2]$ as a function of the iteration
    count both for the SGD and ADAM algorithms. Both parameters
    $\theta_i, \theta_2$ tend to their optimal value (zero).}
  \label{fig:sgd}
\end{figure}

As an example we consider the problem of minimizing $\mathbb
E_\theta[x_1^2 + x_2^2]$ (i.e. 
finding the values for $\theta$ that make $\mathbb
E_\theta[x_1^2+x_2^2]$ minimum). 
We have implemented two flavours of Stochastic Gradient Descent (SGD): the first -basic- one, having a constant learning rate, and the second one being the well-known
ADAM algorithm \cite{kingma2017adam}. It is worth noting at this
point that as a general concept, SGD implies a stochastic (but
unbiased) evaluation of the gradients of the objective function at
every iteration. While in typical applications in the ML community,
where the task is to fit some dataset, this is done by evaluating the
gradients at different random batches of the data, the present example
is different in that no data is involved. In this case, every
iteration of the SGD evaluates the gradients on the different Monte
Carlo samples used to approximate the objective function  $\mathbb
E_\theta[x_1^2 + x_2^2]$.  

Here we consider a simple implementation of the Metropolis Hastings algorithm in order to
first produce the samples $\{x^{\alpha}\}_{\alpha=1}^N \sim p_\theta(x)$. 
Second, we determine the reweighted expectation value truncated at
first order
\begin{equation}
  \frac{\sum w(x^{\alpha};\tilde\theta) \left[ (x^{\alpha}_1)^2+ (x_2^{\alpha})^2 \right]}{\sum w(x^{\alpha};\tilde \theta)} \approx \bar O + \bar O_i \epsilon_i\,,  
  \qquad \left( w(x^{\alpha};\theta) = e^{S(x^{\alpha};\theta) - S(x^{\alpha};\tilde \theta)}  \right)\,,
\end{equation}
where $\tilde \theta_i =  \theta_i + \epsilon_i$. 
This quantity gives an stochastic estimate of the function value
\begin{equation}
  \bar O = \frac{1}{N}\sum_{i=1}^N [x_1^{\alpha}]^2 + [x_2^{\alpha}]^2\,,
\end{equation}
and its derivatives
\begin{equation}
 \bar O_i \approx \frac{\partial \mathbb E_\theta[x_1^2+x_2^2]}{\partial \theta_i}\,.
\end{equation}

Figure~\ref{fig:sgd} shows the result of the optimization process. 
As the iteration count increases the function is driven to its minima,
while the values of the parameters approach the optimal values
$\theta_1^{\rm opt} = \theta_2^{\rm opt} = 0$. 

It is worth mentioning that in this particular example only $1000$ samples
were used at each step to estimate the loss function and its
derivatives. 
If one decides to use a larger number of samples (say $10^5$), the
value of the parameter $\theta_2$ is determined with a much better precision. 
Note that the direction associated
with $\theta_2$ is much flatter, and therefore its value affects much
less value of the loss function.

\subsection{An application in Bayesian inference}
\label{sec:bayesian}
The purpose of statistical inference is to determine properties of the
underlying statistical distribution of a dataset
$D=\{x_{i},y_{i}\}_{i=1}^{N}$. In many
  cases, the independent variables $x_i$ are fixed, and all the
  stochasticity is captured by the dependent variables $y_i$. As
  such,  
the data is assumed to be sampled from a certain model, specified by
the \textit{likelihood}, 
$p(y|x,\phi)$, which depends on a set of parameters $\phi$.
The Bayesian paradigm attributes a level of confidence to the model by
introducing the \textit{prior} 
$p_{\theta}(\phi)$, \textit{i.e.} an a priori distribution of the
models parameters, where in this context $\theta$ play the role of the
hyper-parameters specifying the prior. Following Bayes' rule, the
\textit{posterior} distribution $p_{\theta}(\phi|D)$ is computed
as\footnote{The normalization factor, 
  $p_{\theta}(D)$, called the evidence, or marginal likelihood, is $\phi$-independent
  and represents the probability distribution of the observed data, given the model.}:
\begin{equation}
  \label{eq:bayes}
  p_{\theta}(\phi|D) \propto p(D|\phi) p_{\theta}(\phi)~.
\end{equation}
The likelihood of the whole dataset, $p(D|\phi)$, is computed assuming independent data points following a Gaussian distribution:
\begin{equation}
  p(D|\phi) = \prod_{i=1}^{N}\mathcal N(y_{i}|f(x_{i};\phi),\sigma_{i})\,,
  \label{eq:likelihood}
\end{equation}
where $\sigma_i$ are the  uncertainties of the corresponding observations $y_i$ (and assumed here to be given), while the mean of the Gaussian is given by $f(x_i;\phi)$. 
From a practical standpoint, in addition to the normalization being,
in general, unknown, the usual complexity of the posterior
distribution makes this possibly highly dimensional integral difficult
to compute. The use of Monte Carlo techniques, in particular of the
HMC, is typical in this context.
We focus below on two types of predictions: 1) The variance of the
model parameters $\delta\phi_j^2 = \mathbb{E}_{p_\theta}[\phi_j^2] -
(\mathbb{E}_{p_\theta}[\phi_j])^2$, where $j=1,...,d$, being $d$ the
dimension of $\phi$, and 2) the variance of the output mean $\delta
f_t^2 = \mathbb{E}_{p_\theta}[f_t^2] -
(\mathbb{E}_{p_\theta}[f_t])^2$, where $f_t$ is a shorthand notation
for the output mean $f(x_t;\phi)$, evaluated at a new ``test''
datapoint $x_t$ \footnote{Note that $E_{p_\theta}[f_t]$ is analogous
  to the so-called ``predictive distribution'' of Bayesian inference,
  however here we focus on the expected value of the prediction mean,
  instead of the expected value of the likelihood of $y(x_t)$
  itself.}.

We are interested in studying the dependence of these quantities on the choice of
hyperparameters $\theta$ that characterize the prior distributions. In
particular we will consider the case of Gaussian priors, and determine
the dependence of our predictions with the width of this Gaussian.

\subsubsection{Model and data set}

\begin{figure}
  \centering
  \includegraphics[width=0.8\textwidth]{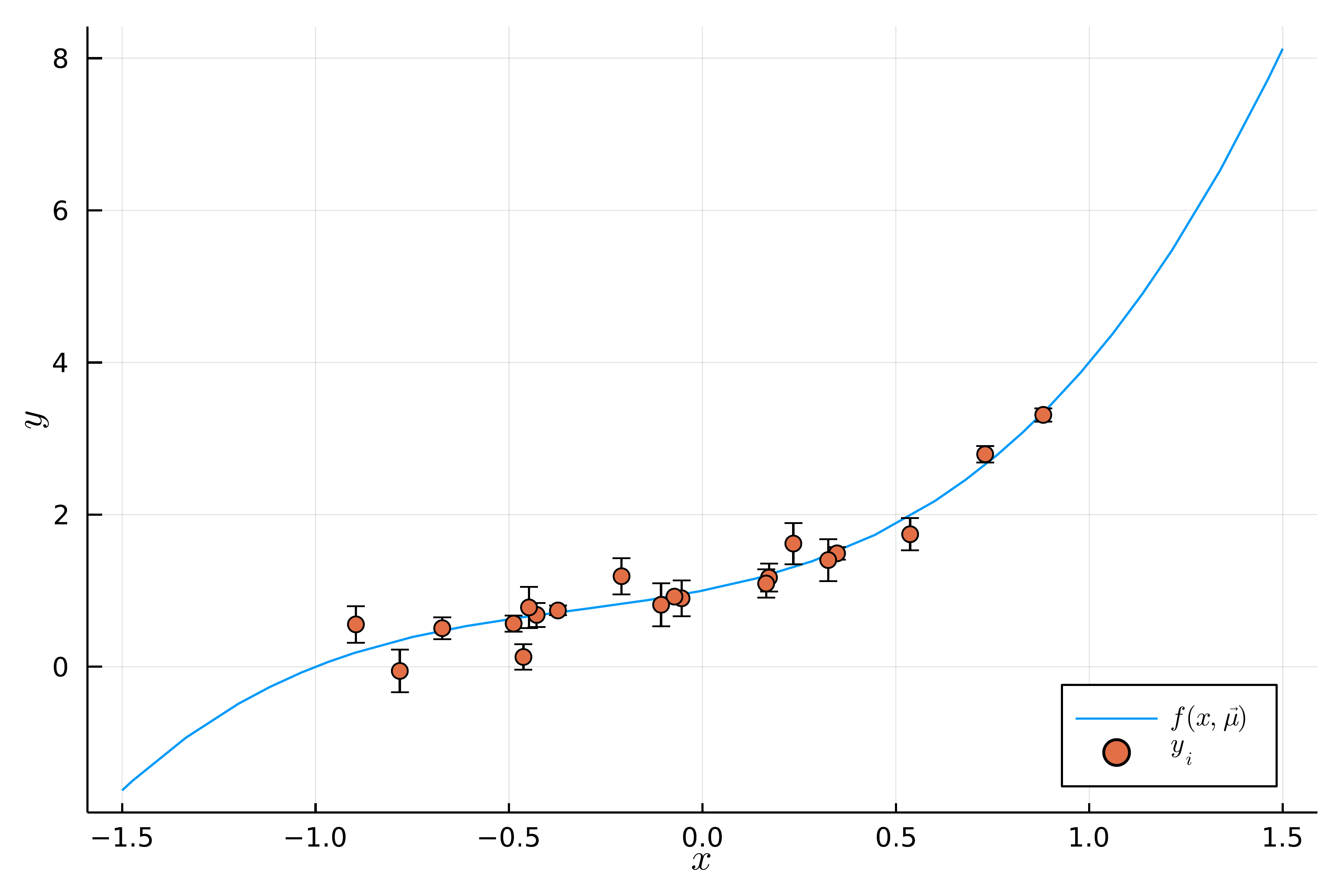}
  \caption{Third order polynomial $y(x)=1+x+x^{2}+x^{3}$ and the randomly sampled dataset $D$.}
  \label{fig:dataset}
\end{figure}

We generate a synthetic dataset (cf. Figure \ref{fig:dataset}) by defining the points on an irregular grid in the range $x_i\in[-1.0;1.0]$, such that
\begin{equation}
  y_i = f(x_i;\phi_{\rm true}) + \sigma_i\epsilon~,
\end{equation}
where the mean is a 3rd degree polynomial, $f(x;\phi)=\phi_0+\phi_1x + \phi_2x^2 + \phi_3x^3$, with $\phi_{\rm true} = (1,1,1,1)$; $\epsilon\sim{\cal N}(0,1)$ is sampled from a standard Gaussian, and we consider a heteroscedastic dataset by defining a noise $\sigma_i$ dependent on $x_i$. We adopt the same model in order to  make inference on the parameters $\phi$. 

The prior distribution is also chosen as a Gaussian, $\phi\sim {\cal N}(\mu_p,\sigma_p)$. 
For simplicity we choose the priors centered on the ``correct'' values
of the model (i.e. $\mu_p =\phi_{\rm true}$), while we keep the width 
$\sigma_p$ as a hyperparameter to study the dependence on\footnote{This is a simplified setup for the sake of illustration, given the methodological scope of this work. Nonetheless, it is straightforward to apply the method to the situation where we are interested in studying the dependence on both parameters $\mu_p$ and $\sigma_p$ simultaneously, or in general on the joint set of hyperparameters of the model. }.

For any choice of the prior width $\sigma_p$ we can obtain a prediction by
generating $N$ samples $\{\phi^{(\alpha)}\}^N_{\alpha=1}$ according to the distribution
$p_{\theta}(\phi|D)$ computed from \cref{eq:bayes}.  
  
\subsubsection{Reweighting approach}
\label{sec:bayesianhmc}

The reweighting method takes $N$ samples
$\{\phi_{i}^{({\alpha})}\}_{\alpha=1}^{N}$ obtained at
$\sigma_{p}=\sigma_{p}^{*}$ and computes the reweighted average using
$\tilde\sigma_{p}=\sigma_{p}^{*}+\epsilon$ in \cref{eq:rw}.  

For each sample $\phi^{(\alpha)}$, the reweighting factor becomes a polynomial expansion in
$(\sigma-\sigma_{p}^{*})$  
\begin{equation}
  \label{eq:rw bi}
  \tilde w_{\alpha}(\epsilon) = \frac{p_{\mu,\sigma_{p}^{*}+\epsilon}(\phi_{\alpha}|D)}{p_{\mu,\sigma_{p}^{*}}(\phi_{\alpha}|D)}.
\end{equation}
Notice that the zeroth order of \cref{eq:rw bi} is one, such that the zeroth order result corresponds to the usual Monte Carlo point estimate for $\delta\phi_{0}(\sigma_{p}^{*})$.

In order to generate these samples, we used the standard HMC algorithm. 
The equations of motion are
\begin{align}
  &H_{\theta}(\phi,\pi) = \frac{\pi^{2}}{2} - \log(p_{\theta}(\phi|D)),\\
	&\dot\phi_{j} = \pi_{j},\\
	&\dot \pi_{j} = - \frac{1}{\sigma_{p}^{2}}(\phi_{j} - (\mu_p)_j) + \sum_{i=0}^{N}\frac{1}{\sigma_{i}^{2}}\left( y_{i} - f(x_{i},\phi) \right)(x_{i})^{j},
\end{align}
where $\pi=\{\pi_{0},\pi_{1},\pi_{2},\pi_{3}\}$ are the momenta conjugated to $\phi$.
Note that all $\phi$-independent terms can be dropped from the
equations of motion, namely the normalization of $p_{\theta}(\phi|D)$
is not needed.
The eom were solved numerically using a fourth-order symplectic
integrator \cite{OMELYAN2003272} providing a high acceptance rate in
the Metropolis-Hastings step even with a coarse integration.  

The chosen integration step-size was $\varepsilon = 0.001$, while the
trajectory length was uniformly sampled in the interval $[0,100]\times
\varepsilon$\footnote{Due to the quadratic form of the Hamiltonian,
  the phase space of this system is cyclic. The algorithm is ergodic
  only if the trajectory length is
  randomized \cite{RHMC2017}.}.



\subsubsection{Hamiltonian perturbative expansion}

Following the procedure in \cref{sec:nspt}, the Monte Carlo samples
$\{(\tilde\phi_{j})^{\alpha}\}_{\alpha=1}^{N},~j=0,1,2,3$ were
obtained with the modified HMC algorithm for some values of
$\sigma_{p}^{*}$. 
We used the same parameters for the HMC as described in the previous
section. In particular our acceptances were so close to 100\% that any
bias due to the missing accept/reject step is negligible. 
We checked this hypothesis by further performing another simulation
with a coarser value of the integration step and finding completely
compatible results.

\subsubsection{Results}

\begin{table}[t]
  \centering
  \caption{Results for the expansion coefficients of the variance,
    ${\delta\phi^{2}_{j,n}}$ for $\sigma_{p}^{*}=0.3$
    from the reweighting and hamiltonian expansion.}
\scalebox{0.9}{
  \begin{tabular}{cccccccc}
     & & \multicolumn{6}{c}{$n$} \\\cmidrule{3-8}
     & & 0 & 1 & 2 & 3 & 4 & 5  \\
    \midrule
\multirow{2}{*}{$\delta\phi^{2}_{0,n}$} & RW &    0.00014705(86) &    0.0001384(63) &    -0.000248(29) &     0.000367(62) &     -0.00071(51) &      -0.0003(12)  \\
                  & HAD &   0.00014705(86) &    0.0001365(34) &   -0.0002850(60) &     0.000311(20) &     0.000178(77) &     -0.00115(26)  \\

    \midrule
\multirow{2}{*}{$\delta\phi^{2}_{1,n}$} & RW &     0.01099(15) &       0.0285(12) &      -0.0450(58) &        0.032(13) &         0.04(10) &        -0.61(25)  \\
                  & HAD &     0.01099(15) &      0.02787(69) &      -0.0518(11) &       0.0248(38) &        0.189(16) &       -0.700(46)  \\
    \midrule
\multirow{2}{*}{$\delta\phi^{2}_{2,n}$} & RW &      0.008938(74) &      0.00830(28) &      -0.0283(10) &       0.0850(39) &       -0.234(18) &        0.603(78) \\
                  & HAD &     0.008938(74) &      0.00817(15) &     -0.02789(42) &       0.0849(13) &      -0.2505(44) &        0.726(15)  \\
    \midrule
\multirow{2}{*}{$\delta\phi^{2}_{3,n}$} & RW &     0.03617(59) &      0.1205(51) &       -0.182(24) &        0.050(61) &         0.63(42) &         -4.0(12)  \\
                  & HAD &     0.03617(59) &       0.1177(30) &      -0.2052(42) &        0.020(16) &        1.132(66) &        -4.02(19)  \\
    \bottomrule
    \label{tab:variance phi0}
  \end{tabular}
  }
\end{table}

Here we compare the predictions for the average model parameters
$\phi$ and their dependence on the prior width $\sigma$. In particular
we focus on the variance of the model parameters $\delta\phi^{2}_j$,
since these are the quantities most sentitive to the prior width (i.e. 
very thin priors result in small variance for the model
parameters). We have fixed $\sigma^* = 0.3$, but similar conclusions
are obtained for other values.  

The results of the Monte Carlo average for $\delta\tilde\phi^{2}_i$ and
its derivatives with respect to $\sigma$ are
shown in \cref{tab:variance phi0}. 
Results labeled ``RW'' use the reweighting method, while results
labeled ``HAD'' use the Hamiltonian approach. 

\begin{figure}[htb!]
  \centering
  \begin{subfigure}{0.45\textwidth}
    \includegraphics[width=\textwidth]{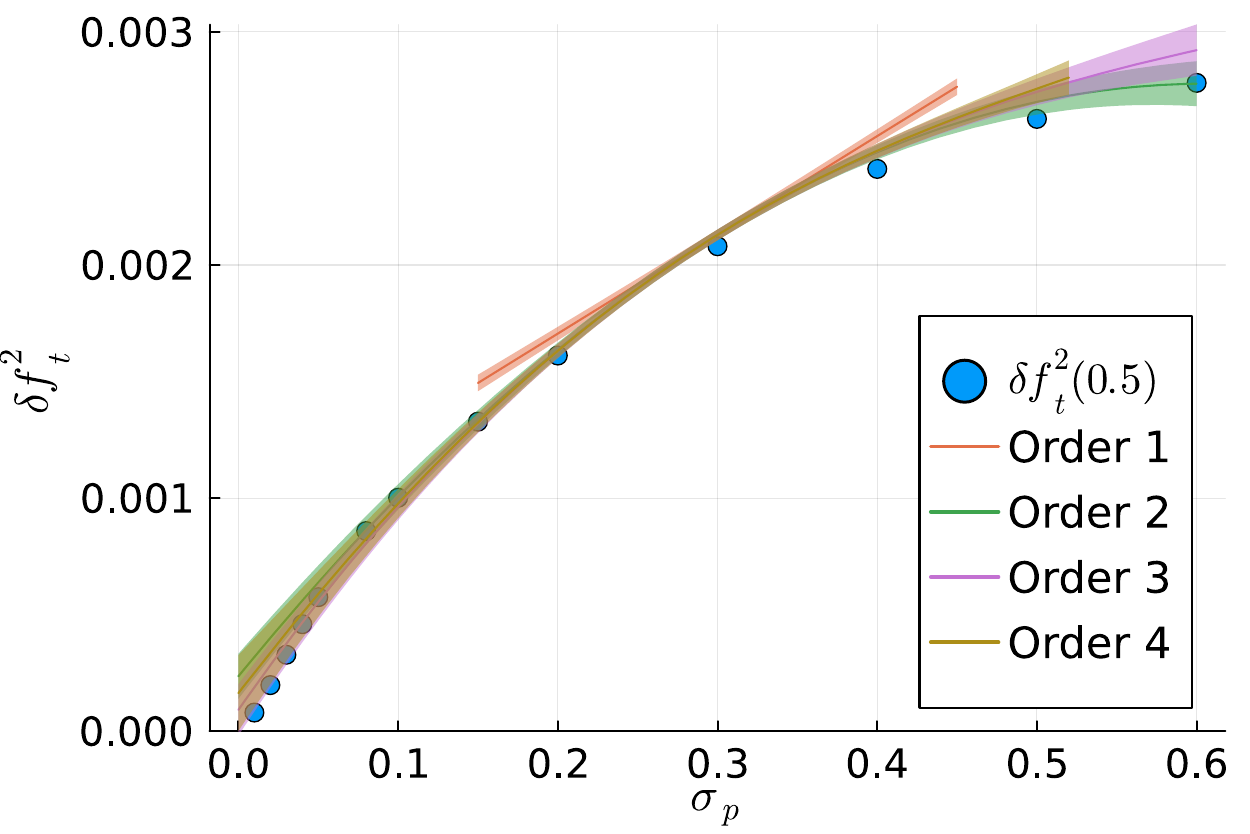}
    \caption{Reweighting approach}
  \end{subfigure}
  \begin{subfigure}{0.45\textwidth}
    \includegraphics[width=\textwidth]{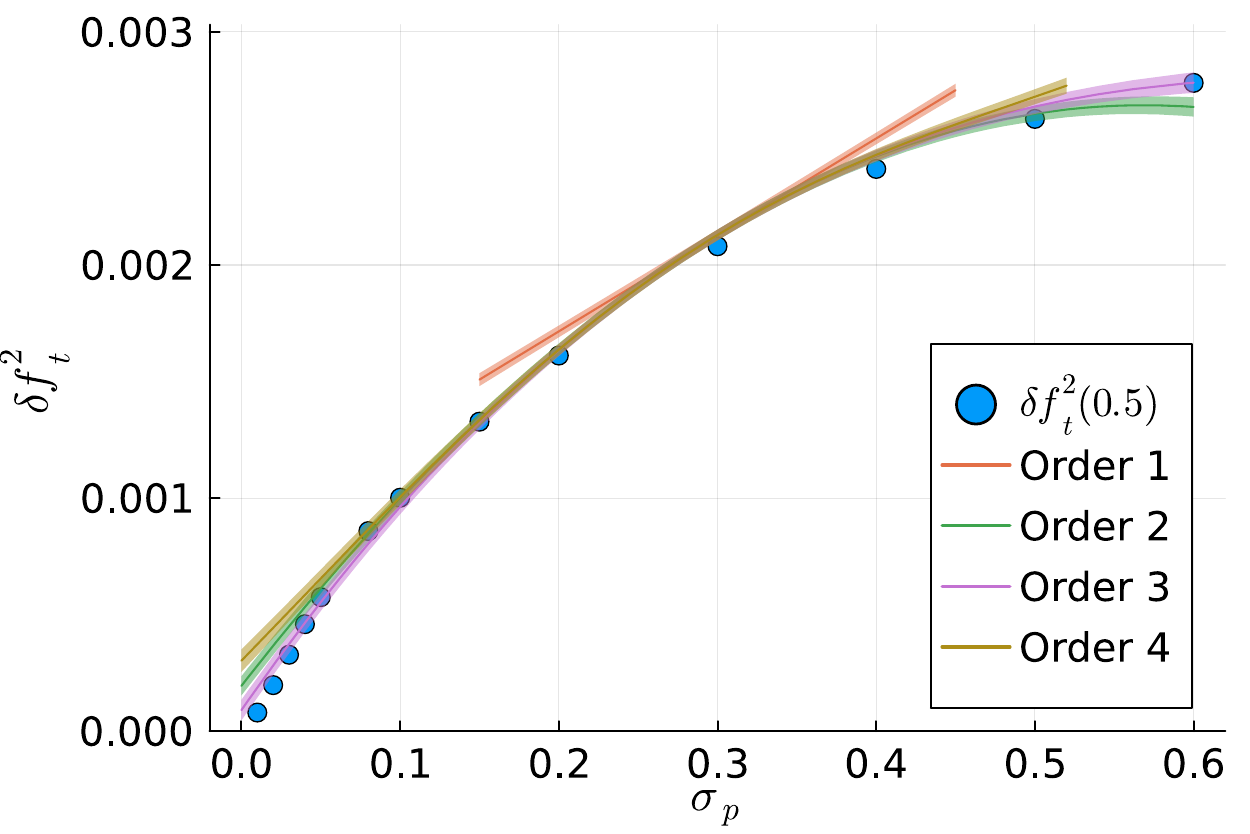}
    \caption{Hamiltonian approach}
  \end{subfigure}
  \caption{Variance of the prediction $\mathbb
    E_{p_{\theta}}[f(x_t,\phi)]$ for a new point $x_t=0.5$ as a
    function of $\sigma_{p}$.  
    The blue points represent simulations at different values of
    $\sigma_{p}$. 
    The error bands represent the evaluation of the Taylor series (up
    to $\order{(\sigma-\sigma_{p}^{*})^{4}}$ with
    $\sigma_{p}^{*}=0.3$) where derivatives are  
    obtained with the Hamiltonian based approach (right)
    or the reweighting approach (left). 
  The Hamiltonian approach gives more precise results. 
  }
  \label{fig:ypred}
\end{figure}

It is obvious that results using the Hamiltonian approach are more
precise:
the uncertainties in the derivatives, $\delta\phi^{2}_{i,n},n\neq 0$, are
smaller for the Hamiltonian approach, despite the statistics being the
same. The difference is larger for higher order derivatives: the
approach based on reweighting struggles to get a signal for the fourth
and fifth derivatives, while the Hamiltonian approach is able to
obtain even the fifth derivative with a few percent precision. 
This fits our expectations (see section~\ref{sec:hamilt-appr-repar}). 
\noindent\newline\newline
On the other hand, for our second quantity of analysis $\delta f_t^2$ (i.e. the variance of the prediction mean), Figure~\ref{fig:ypred} shows the results of the dependence on $\sigma_p$, where we have fixed $x_t=0.5$.

The Hamiltonian approach gives visually results with a reduced variance,
similar to the results presented in table~\ref{tab:variance phi0}.

\section{A case study in lattice field theory}
\label{sec:lamphi4}

In this section we explore in detail some applications in lattice field theory. 
We will use as model the $\lambda-\phi^4$ theory in 4 space-time dimensions. 
In the continuum the Euclidean action of this theory is given by
\begin{equation}
  S(\phi; m, \lambda) = \int {\rm d} x^4 \left\{  \frac{1}{2}(\partial_\mu\phi)^2 + \frac{m^2}{2}\phi^2 + \lambda\phi^4 \right\}\,.
\end{equation}
The discretized version of this action is
\begin{equation}
  \label{eq:Slatt}
  S_{\rm latt}(\hat \phi;\hat m,\lambda) = \sum_x \left\{  \frac{1}{2}\sum_\mu[\hat\phi(x+\hat\mu) - \hat\phi(x)]^2 + \frac{\hat m^2}{2}\hat\phi^2(x) + \lambda\hat\phi^4(x) \right\}  
\end{equation}
where dimension-full quantities have been scaled with appropriate
powers of the lattice spacing $a$ in order to render all quantities
dimensionless
\begin{eqnarray}
  \hat \phi &=& a\phi\,,\\
  \hat m &=& am\,.
\end{eqnarray}

In field theory one is interested in correlation functions, given as
expectation values over the Euclidean partition function
\begin{equation}
  \mathcal Z = \int \prod_x{\rm d} \hat\phi(x) e^{-S_{\rm latt}(\hat\phi;\hat m, \lambda)}\,.
\end{equation}
These expectation values depend on the parameters $\hat m,\lambda$. 
The methods described in sections~\ref{sec:reweighting}
and~\ref{sec:nspt} can be applied to determine the dependence of
correlation functions with these parameters.
In particular we note that for non-negative $\hat m^2$ the potential
given by the action of eq.~(\ref{eq:Slatt}) is convex, guaranteeing
the convergence of the Hamiltonian perturbative expansion (see section~\ref{sec:impl-conv}). 

We have performed simulations on a $L^4$ lattice with $L/a=32,48$ for several
values of the parameter $\lambda$ and $\hat m^2 = 0.05$. 
As example observables, we will consider some simple local quantities:
$\langle \hat\phi^2(x) \rangle, \langle \hat\phi^4(x) \rangle$, as well as the
action density
\begin{equation}
  \label{eq:act}
  \langle \hat s(x) \rangle =  \frac{1}{2}\langle [\hat\phi(x+\hat\mu) - \hat\phi(x)]^2\rangle + \frac{\hat m^2}{2}\langle \hat\phi^2(x)\rangle + \lambda\langle \hat\phi^4(x)  \rangle\,.
\end{equation}
Since we perform our simulations with periodic boundary conditions,
invariance under translations ensures that these local expectation
values are independent on the point measured $x$. 
In order to get better precision we perform volume averages for the
estimations (\textit{e.g.}
$ \langle \tilde s \rangle = \left(\frac{a}{L}\right)^4 \sum_x \langle \tilde s(x) \rangle$).

Table~\ref{tab:lp4} shows the results of both the Hamiltonian and the
reweighting approach applied to the determination of the derivatives
$\partial/\partial m^2$, $\partial/\partial \lambda$,
$\partial^2/\partial m^2\partial\lambda$ of the observables $\langle
\phi^2 \rangle,\langle \phi^4 \rangle, \langle \hat s \rangle$ (see eq.~(\ref{eq:act})).

\begin{landscape}
\begin{table}
  \centering
  \begin{tabular}{lllllllll}
    \toprule
    &&&\multicolumn{6}{c}{$\lambda$} \\
    \cline{4-9}
    &&&0.0&0.1&0.2&0.3&0.4&0.5\\
    \midrule
    \multirow{6}*{$\langle \phi^2 \rangle$}&\multirow{2}*{$\partial_{\hat m^2}$}
     & RW &     -0.0428(20)&     -0.0328(14)&     -0.0270(13)&     -0.0241(12)&     -0.0220(11)&    -0.01974(91) \\
    && HAD &   -0.042526(41)&   -0.030880(14)&   -0.026273(10)&  -0.0233672(82)&  -0.0212721(72)&  -0.0196387(60) \\
    \cline{2-9}
    &\multirow{2}*{$\partial_{\lambda}$}
     & RW &     -0.0779(22)&    -0.05227(94)&    -0.04370(89)&    -0.03534(61)&    -0.03169(50)&    -0.02754(49) \\
    && HAD &   -0.077816(79)&   -0.052499(24)&   -0.042218(19)&   -0.035830(14)&   -0.031323(11)&  -0.0278909(93) \\
    \cline{2-9}
    &\multirow{2}*{$\partial^2_{\hat m^2,\lambda}$}
     & RW &        0.43(43)&        0.03(16)&        0.16(14)&       -0.10(11)&       0.116(77)&      -0.024(69) \\
    && HAD &      0.2733(22)&    0.061593(99)&    0.035082(69)&    0.024240(42)&    0.018263(31)&    0.014553(31) \\
    \midrule
    \multirow{6}*{$\langle \phi^4 \rangle$}&\multirow{2}*{$\partial_{\hat m^2}$}
     & RW &     -0.0391(20)&     -0.0272(12)&     -0.0223(11)&    -0.01809(90)&    -0.01645(86)&    -0.01398(70) \\
    && HAD &   -0.038919(39)&   -0.026247(13)&  -0.0211084(97)&  -0.0179118(73)&  -0.0156615(62)&  -0.0139464(46) \\
    \cline{2-9}
    &\multirow{2}*{$\partial_{\lambda}$}
     & RW &     -0.0844(24)&     -0.0539(11)&    -0.04281(92)&    -0.03340(54)&    -0.02850(43)&    -0.02428(41) \\
    && HAD &   -0.084330(78)&   -0.054229(26)&   -0.041514(19)&   -0.033715(14)&   -0.028357(11)&  -0.0243679(73) \\
    \cline{2-9}
    &\multirow{2}*{$\partial^2_{\hat m^2,\lambda}$}
     & RW &        0.41(44)&        0.01(16)&        0.11(14)&      -0.083(90)&       0.089(61)&      -0.000(56) \\
    && HAD &      0.2848(21)&    0.068858(94)&    0.038917(70)&    0.026391(38)&    0.019393(29)&    0.015080(25) \\
    \midrule
    \multirow{6}*{$\langle s \rangle$}&\multirow{2}*{$\partial_{\hat m^2}$}
     & RW &         -0.0025(42)&          -0.0006(34)&      0.0027(35)&      0.0028(36)&      0.0057(32)&      0.0063(30) \\
    && HAD &   -0.000003(22)&    0.002623(16)&    0.004218(20)&    0.005397(14)&    0.006265(16)&    0.006989(15) \\
    \cline{2-9}
    &\multirow{2}*{$\partial_{\lambda}$}
     & RW &         -0.0686(48)&          -0.0567(26)&     -0.0538(25)&     -0.0447(23)&     -0.0400(17)&     -0.0343(17) \\
    && HAD &   -0.069774(49)&   -0.057738(34)&   -0.050128(40)&   -0.044530(27)&   -0.040250(27)&   -0.036721(26) \\
    \cline{2-9}
    &\multirow{2}*{$\partial^2_{\hat m^2,\lambda}$}
     & RW &             1.1(1.0)&        -0.16(39)&        0.36(43)&       -0.43(32)&        0.16(26)&       -0.15(21) \\
    && HAD &    0.038864(96)&    0.019197(66)&    0.013405(69)&    0.010126(50)&    0.007860(47)&    0.006407(48) \\
    \midrule
    \bottomrule
  \end{tabular}
  \caption{Derivatives with respect to $\hat m^2$ ($\partial_{\hat
      m^2}$), $\lambda$ ($\partial_\lambda$) and the cross derivative 
  ($\partial^2_{\hat m^2,\lambda}$) of different observables. 
Note that the action density $s$ has an explicit dependence on
$\hat m^2,\lambda$ (see Eq.~(\ref{eq:act})). 
All simulations are performed on a $L^4$ lattice with $L/a=32$ and
$\hat m^2 = 0.05$.}
  \label{tab:lp4}
\end{table}
\end{landscape}


It is apparent that results obtained with the
Hamiltonian approach are much more precise (both results use exactly
the same statistics), usually about 100 times more precise for the
first derivatives. This difference in precision is even more clear for
the higher orders: the signal for the cross derivative $\partial^2_{\hat
  m^2,\lambda}$ is completely lost in the reweighting approach,
whereas the Hamiltonian approach determines its value with a precision
better than a 1\%. 
This can
be understood from a general point of 
view by noting that the reweighting approach requires to evaluate the
so called \emph{disconnected} contributions. 
  For example, the leading order derivative $\partial_{\hat m^2}
  \langle \phi^2 \rangle$ as determined by the reweighting approach is
  given by
  \begin{equation}
    \partial_{\hat m^2}
    \langle \hat \phi^2 \rangle = \langle \hat \phi^2 (\partial_{\hat m^2}S_{\rm latt})\rangle
    - \langle \hat \phi^2 \rangle \langle\partial_{\hat m^2}S_{\rm latt}\rangle
  \end{equation}
These disconnected contributions are known to suffer from a large
variance. As explained in section~\ref{sec:hamilt-appr-repar}
  the Hamiltonian approach completely avoids estimating such
  disconnected contributions. 
  The Hamiltonian approach implements an exact version
  of the reparametrization trick, where the field variables $\tilde
  \phi(x)$ (and its dependence with the relevant parameters) are
  determined to any order so that all terms in the Taylor series are
  determined \emph{as connected contributions}.

We conclude that it is the absence of disconnected terms in the
Hamiltonian approach what lies at the heart of the differences in variances
between both approaches.


\section{Conclusions}
\label{sc:conclusions}

The tools of automatic differentiation represent a cornerstone in
modern optimization algorithms and many machine learning applications. 
The extension of these techniques to functions evaluated using Monte
Carlo processes is non trivial. In these cases the underlying
probability distribution depends on some parameters, and Monte Carlo
techniques are used to draw samples for specific values of the parameters. 
The dependence of the samples with the parameter values is difficult
to determine.
Nonetheless, there are many applications for these techniques. 
In this work we have considered several of them, from optimizations of
expectation values, to the study of the dependence of Bayesian
predictions with respect to prior parameters or applications in
lattice field theory.

We have presented two different approaches to the
determination of Taylor series of quantities
estimated via Monte Carlo sampling.  
The first approach is based on reweighting and can be considered a
generalization of the score function estimator, valid for derivatives
of arbitrary orders, and unnormalized probability
distribution functions.
The second approach is based on Hamiltonian methods to sampling
(HMC being the most popular option), and produces samples  that carry
the information of the dependence on the 
action parameters. The convergence of the stochastic process in this
last approach is not always guaranteed, but we have provided
sufficient conditions for the convergence.

We have shown some applications of these methods. 
First, in the context of optimization, we have applied the stochastic
gradient descent to find the optimal parameters of some expectation
value (see section~\ref{sec:opt}). 
Second, in Bayesian inference we have shown how these methods can be
used to estimate the dependence on Bayesian predictions on
the ``hyperparameters'' that describe the prior distribution (cf section~\ref{sec:bayesian}).

Finally, in the context of Lattice Field Theory, we have studied in
detail the case of $\lambda-\phi^4$ in four space-time dimensions. 
The dependence of observables with respect to the parameters of the
action (the bare mass in lattice units $\hat m$ and the bare coupling
$\lambda$) can be accurately determined using these techniques. 

A detailed comparison of both methods shows that results obtained with
the Hamiltonian approach are much more precise. 
We have argued that the Hamiltonian approach can be seen as a change
of variables from the reweighing formula where the reweighing factors
are constant: 
In the Hamiltonian approach all dependence with respect to the
parameters is present in the samples. 
The absence of disconnected contributions (in the lattice jargon)
makes the variance of Taylor series computed with the Hamiltonian
approach much more precise. For example, our study on $\lambda-\phi^4$
shows that for the same statistics one gets results 100 times more precise. 

The Hamiltonian approach has his own drawbacks. 
On one hand the convergence of the stochastic process is not guaranteed. 
In particular for the case of Lattice QCD, that is formulated in terms
of compact variables, the convergence cannot be guaranteed. 
On the other hand, the method cannot be applied to samples that have
already been generated, unlike the reweighting method.

Nevertheless the investigations of this work open the door to an
interesting possibility: that one can find change of variables that
eliminate (or significantly reduce) the disconnected contributions of
the reweighing approach. 
Machine Learning techniques, in particular the tools related with
normalizing flows, potentially can provide a significant gain in the
computation of derivatives of expectation values.


\section*{Acknowledgments}
\addcontentsline{toc}{section}{Acknowledgments}
The  authors are gratefult to A. 
Patella for the many discussions on the early stages of the work presented here, as well as A. Dimitriou, D. Hernandez-Lobato and S. Rodriguez-Santana.
AR and GT acknowledge financial support from the Generalitat
Valenciana (CIDEGENT/2019/040). Similarly, BZ akcnowledges the support from CIDEGENT/2020/055. 
The authors gratefully acknowledge as well the support from the Ministerio de
Ciencia e Innovacion (PID2020-113644GB-I00) and computer resources at
Artemisa, funded by the European Union ERDF and Comunitat Valenciana
as well as the technical support provided by the Instituto de Física
Corpuscular, IFIC (CSIC-UV). The authors acknowledge the financial support from the MCIN with funding from the European Union NextGenerationEU (PRTR-C17.I01) and Generalitat Valenciana. Project ``ARTEMISA'', ref. ASFAE/2022/024.

\appendix

\clearpage
\addcontentsline{toc}{section}{References}
\bibliography{refs}

\end{document}